# MUON CAPTURE IN THE FRONT END OF THE IDS NEUTRINO FACTORY *

D. Neuffer, Fermilab, Batavia, IL 60510, USA
M. Martini, G. Prior      ve, Suisse
C. Rogers, RAL ASTeC, Chilton, Didcot UK
C. Yoshikawa, Muons, Inc., Batavia IL 60510, USA

*Abstract*

We discuss the design of the muon capture front end of the neutrino factory International Design Study. In the front end, a proton bunch on a target creates secondary pions that drift into a capture transport channel, decaying into muons. A sequence of rf cavities forms the resulting muon beams into strings of bunches of differing energies, aligns the bunches to (nearly) equal central energies, and initiates ionization cooling. The muons are then accelerated to high energy where their decays provide neutrino beams. For the International Design Study (IDS), a baseline design must be developed and optimized for an engineering and cost study. We present a baseline design that can be used to establish the scope of a future neutrino Factory facility.

## INTRODUCTION

The goal of the IDS-Neutrino Factory is to deliver a reference design report by 2012 in which the physics requirements are specified and the accelerator and detector systems are defined, with an estimate of the required costs.[1] It consists of:

- a proton source with a baseline intensity goal of 4MW beam power (50Hz, ~10GeV protons, ~2ns bunches. (~5×$10^{13}$ p/ bunch),
- a target, capture and cooling section that produces $\pi$'s that decay into $\mu$'s and captures them into a small number of bunches.
- an accelerator that takes the $\mu$'s to 25 (or 50) GeV and inserts them into storage rings. $\mu$ decay in the straight sections provides high-energy $\nu$ beams for:
- ~100 kton $\nu$-detectors at 4000-7500km baselines with sufficient resolution to identify $\nu$-interactions.

The goal is > $10^{21}$ $\nu$'s /beamline/ year in order to obtain precise measurements of $\nu$-oscillation parameters.

The present paper discusses the muon capture and cooling system. In this system we follow ref. [2], and set 201.25MHz as the baseline bunch frequency. The $\pi$'s (and resulting $\mu$'s) are initially produced with broad energy spreads, much larger than the acceptance of any accelerator, and much larger in phase space than a 200MHz rf bucket. In this "front end" system, we capture this large phase space of $\mu$'s into a string of ~200MHz bunches, rotate the bunches to equal energies, and cooled them for acceleration to full energy. The method captures both $\mu^+$'s and $\mu^-$'s simultaneously and can be adapted to feed a $\mu^+$-$\mu^-$ collider.

## IDS BASELINE FRONT END

The baseline front end is shown in Figure 1. ~10 GeV protons are targeted onto a Hg jet target that is encapsulated in a 20 T solenoid. $\pi$'s created from the target are captured as they traverse the 10 m long solenoid, that has a field profile that starts at 20 T and 7.5cm radius at the target and tapers off to ~1.5 T and 30cm radius at the end. This section captures $\pi$'s and $\mu$'s with transverse momenta $p_t < eBr/2 = 0.225$GeV/c, with an adiabatic damping of the transverse momentum.

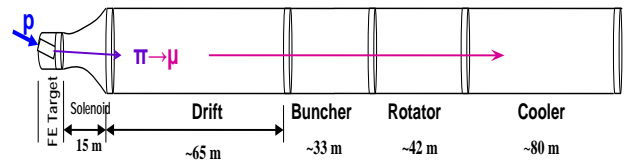

Figure 1: Overview of the front end, consisting of a target solenoid (20 T), a tapered capture solenoid (20 T to 1.5T, 15m long), Drift section (65m), rf Buncher (33 m), an energy-phase Rotator (42m), and a Cooler (~80m).

The taper is followed by a Drift section, where $\pi$'s decay to $\mu$'s, and the bunch lengthens, developing a high-energy "head" and a low-energy "tail". In the Buncher, rf voltages are applied to the beam to form it into a string of bunches of different energies. (fig. 2) This is obtained by requiring that the rf wavelength of the cavity is set to an integer fraction of the c$\tau$ between reference particles:

$$\lambda_{rf}(L) = \frac{L}{N}\left(\frac{1}{\beta_N} - \frac{1}{\beta_0}\right)$$

In the baseline, muons with $p_0 = 232$ MeV/c, and $p_N = 154$ MeV/c with N=10 are used as reference particles. The reference particles (and all intermediate bunch centers) remain at 0-phase throughout the buncher. The rf frequency decreases from 320 to 232 MHz along the 33m Buncher while the rf gradient in cavities increases from 0 to 9 MV/m. In the Rotator, the lower energy reference particle is moved to an accelerating phase as the wavelength separation is also lengthened. (10 → ~10.05) At the end of the Rotator the reference particles are at the same momentum (~232MeV/c) and the rf frequency is matched to 201.25 MHz.. μ's with initial momenta from ~80 to 500 MeV/c have been formed into a train of 201.25 MHz bunches with average momenta of ~232MeV/c and $\delta p_{rms}/p \approx 10\%$. The bunch train is ~60m long with ~40 bunches.

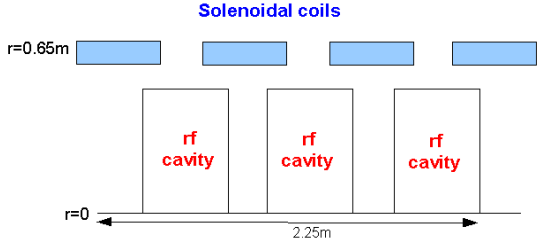

Figure 2: Baseline Layout of RF + magnets in Buncher and Rotator. The rf is in 0.5m cavities with 0.25 drifts, with a 1.5T focusing solenoid field throughout.

The μ's are matched into a cooling section (fig. 3) which consists of rf cavities, LiH absorbers for cooling and alternating solenoids (AS) for focusing. (B oscillates from 2.7 to -2.7T with a 1.5 m period.). The cooling doubles the number of accepted μ's while reducing the rms transverse emittances by a factor of 3. After ~75m of cooling, we find that the system accepts ~0.1μ$^+$/ 10 GeV proton within reference acceptances of $\varepsilon_{L,N}$ <0.2m, $\varepsilon_{t,N}$ <.3cm, which are the acceptances of the downstream acceleration and storage rings. As a bonus, the method simultaneously produces bunch trains of both signs (μ$^+$ and μ$^-$) at equal intensities.

The baseline presented here is a variation of the front end presented in ref. [1] and is ~25% shorter with somewhat smaller magnetic fields and rf gradients. The changes are designed to reduce cost and the risks of large rf gradient requirements.

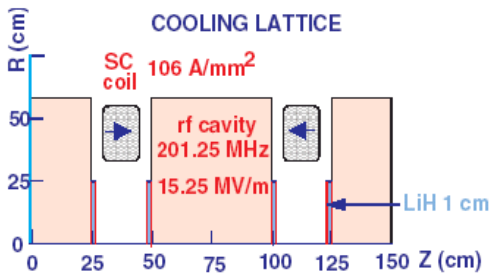

Figure 3: Baseline cooling channel cell layout, with LiH absorbers, rf cavities and focusing coils.

## IMPLEMENTATION FOR THE IDS

For the IDS, the method must be evaluated for practicality and cost. The initial concept had a continuous variation of frequency from cell to cell; an implementation with ~13 separate frequencies in the buncher and ~15 in the rotator is used for a practical implementation. These rf cavities will be grouped into units with matched power supplies, and the configurations will be evaluated.

In a first approximation, the rf cavities are copper pillbox shapes (at 200 MHz, a=0.57m, Q=58000) and are similar to the 200 MHz rf cavities (rounded Cu cylinders with Be windows) built for MICE.[3] From the MICE rf specifications, scaled by the pillbox cavity model, we can estimate the rf requirements of the IDS, and these are summarized in Table 1.

Table 1: Baseline rf requirements

| **Region** | **Number of rf cavities** | **rf frequencies** | **rf gradients, peak power** |
|---|---|---|---|
| Buncher | 37 | 320 to 231.6 MHz, 13 frequencies. | 0-7.5 MV/m, 0.5 to 3.5MW per frequency |
| Rotator | 56 | 230 to 202.3 MHz, 15 freq. | 12 MV/m, ~2.5MW/cavity |
| Cooler (75m) | 100 | 201.25 MHz | 15MV/m, ~4MW / cavity |

The front end will also require significant magnet costs, initially determined from combining the 20 to 1.5 T solenoid r=7.5cm to r=30cm) with a ~150m long 1.5 T solenoid and 75m of AS coils (100coils) for the cooling. This component count, with civil construction, will be used to set an initial cost baseline for the IDS.

Performance estimates and optimizations are being obtained using ICOOL,[4] G4Beamline,[5] and G4MICE[6] particle tracking simulations and the results are consistent. Fig. 4 shows results of a simulation of μ-capture from π's produced by 10000 8-GeVp on target. ~0.09μ/p are within the IDS acceptance. Fig. 5 displays the longitudinal phase space through the system.

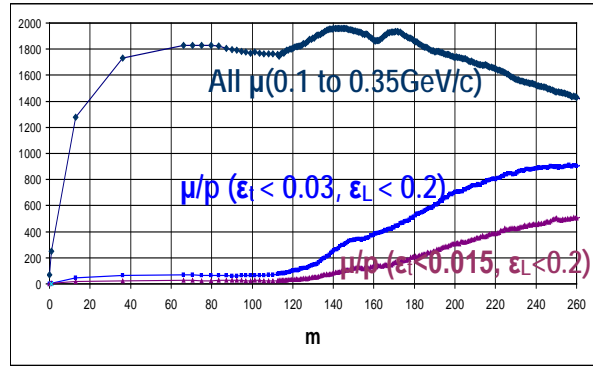

Fig. 4: Captured μ's in the system. μ's are captured in a ~60m long train of λ=1.5m bunches. At z= 160m ~0.04 μ/p are captured in the acceptance. At end of cooling~ 0.09 μ/p have been cooled to within the acceptance. $\varepsilon_{t, rms}$ is cooled from ~0.017 to 0.006m (ICOOL result.)

## RF GRADIENT LIMITATIONS

The μ capture concept requires using relatively high gradient rf fields interleaved with relatively strong solenoidal magnetic fields. In the Buncher, rf gradients of ~7MV/m at ~200MHz within 1.5T solenoids are needed.

The Rotator uses 12MV/m gradients within 1.5T, and the Cooler uses ~15MV/m within AS fields. Recent experiments appear to show that maximum rf gradients are reduced in magnetic fields. Operation of 200 MHz rf within a solenoid will be tested soon.[7] If limitations are found, we have several mitigation strategies that can be used to maintain a practical design. These are described in more detail a separate paper. They include changing the rf cavities by using Be or using open-cell cavities, or using amagnetically-insulated rf cavities, or using magnetically shielding. Gas-filled rf cavities can also be used, since experiments have shown that $H_2$ gas suppresses breakdown in magnetic fields. [8, 9, 10]

*Lower gradient operation*

If limitations exist but are not too stringent, lower gradient operation may provide adequate mitigation. In the present baseline, the buncher rf fields (<~7MV/m) are expected to be relatively safe. In the Rotator, rf at 12MV/m could be reduced to 8MV/m without degradation by increasing the rf occupancy from 2/3 to 1. We have also studied the effects of reduced gradients in simulation, and found that the buncher/rotator acceptance is not greatly reduced by reducing the gradients by as much as a factor of 2.

Reduction of gradients in the Cooler is more difficult, since the rf must compensate for ionization energy loss as well as bunch the beam. (It is possible that the AS field does not limit gradients as much as a constant solenoid.) A 12 MV/m cooler had ~25% less acceptance than the baseline 15MV/m, and much lower gradients could eliminate the gain from cooling.

## VARIATIONS & FUTURE STUDIES

We have presented a baseline design that sets the scale of the IDS front end system. rf R&D may require changes in that baseline, but should not change the scale of the system. Variations that improve performance and/or reduce cost will be considered and developed.

*Research supported by US DOE under contract DE-AC02-07CH11359 and SBIR grant DE-SC-0002739.

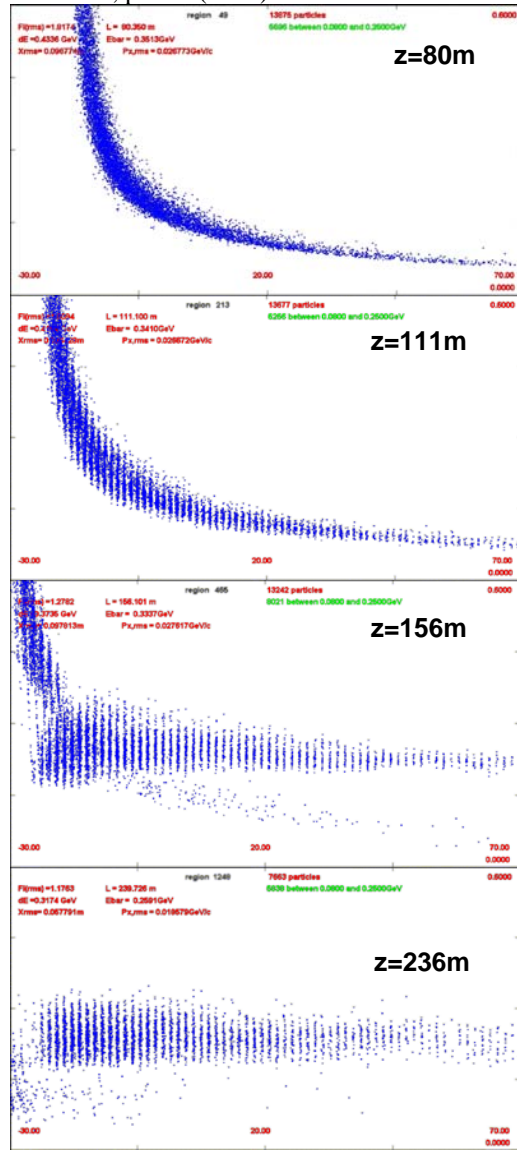

**Fig. 5.** ICOOL simulation results of the buncher and phase rotation. A:µ's as produced at the end of the solenoidal capture + drift.(z=80m) B: µ's at z=111mafter the buncher. C: µ's at z=156m, the end of the rotatorr. The beam has been formed into a string of ~200MHz bunches at ~equal energies. D: At z= 236m after ~80m of cooling. µ's captured within rf buckets are cooled.ϕ-δE rotation; the bunches are aligned into nearly equal energies. In each plot the vertical axis is momentum (0 to 0.6 GeV/c) and the horizontal axis is longitudinal position (-30 to 70m).